\documentclass{llncs}
\usepackage[usenames,dvipsnames]{color}
\usepackage[english]{babel}
\usepackage[latin1]{inputenc}
\usepackage{setspace}
\usepackage{graphicx,times}
\usepackage{amsmath,amssymb}
\usepackage{multicol}
\usepackage[scriptsize,bf]{caption}
\usepackage{float}
\usepackage[absolute,overlay]{textpos}
\usepackage{shadow}
\usepackage{pstricks,pst-grad}
\usepackage{subfig}
\usepackage{sidecap}

\begin{document}
\title{Characterising Steady-State Topologies of \\ SIS Dynamics on Adaptive Networks}
\author{Stefan Wieland\inst{1} \and Tom\'as Aquino\inst{2} \and Andrea Parisi\inst{1} \and Ana Nunes\inst{1}}
\institute{
Centro de F{\'\i}sica da Mat{\'e}ria Condensada and Departamento de F{\'\i}sica, Faculdade de Ci{\^e}ncias da Universidade de Lisboa, P-1649-003 Lisboa, Portugal
\and Centro de Matem\'atica e Aplica\c c\~oes Fundamentais and Departamento de F{\'\i}sica, Faculdade de Ci{\^e}ncias da Universidade de Lisboa, P-1649-003 Lisboa, Portugal}

\maketitle

\begin{abstract}
Disease awareness in epidemiology can be modelled with adaptive contact networks, where the interplay of disease dynamics and network alteration often adds new phases to the standard models (\cite{Gross}, \cite{Shaw}) and, in stochastic simulations, 
lets network topology settle down to a steady state that can be static (in the frozen phase) or 
dynamic (in the endemic phase).  We show for the SIS model that, in the endemic phase, this steady state does not depend on the initial network topology, only on the disease and rewiring parameters and on the  link density of the network, which is conserved. We give an analytic description of the structure of this co-evolving network of infection through its steady-state degree distribution.
\end{abstract}

\section{Elaborating on the Pairwise SIS Model with Rewiring}
Gross et al. (\cite{Gross}) incorporated disease awareness into conventional SIS dynamics by allowing susceptibles to evade infection through retracting links from infected neighbours and rewiring them to randomly selected susceptibles.  Using the moment closure from \cite{Keeling} to obtain a pair approximation model (\emph{PA}) of the process, the time evolution of the fraction of infected \(i\) is coupled to that of the densities \(l_{\mathrm{XY}}\) of links connecting nodes of type \(X\) and \(Y\) (\(\{X,Y\} \in \{S\mathrm{=susceptible},I\mathrm{=infected}\}\)).  Apart from the conserved link density \(l=l_{\mathrm{SS}}+l_{\mathrm{SI}}+l_{\mathrm{II}}\), the model can be parameterised by \(\rho,\omega \in [0,1]\), with rewiring, infection and recovery rates given by \(\omega, \left( 1 - \omega\right)  \rho\) and \(\left(1-\omega\right)\left(1-\rho\right)\), respectively.

\begin{figure}[ht]
\centering
\includegraphics[width=1\textwidth]{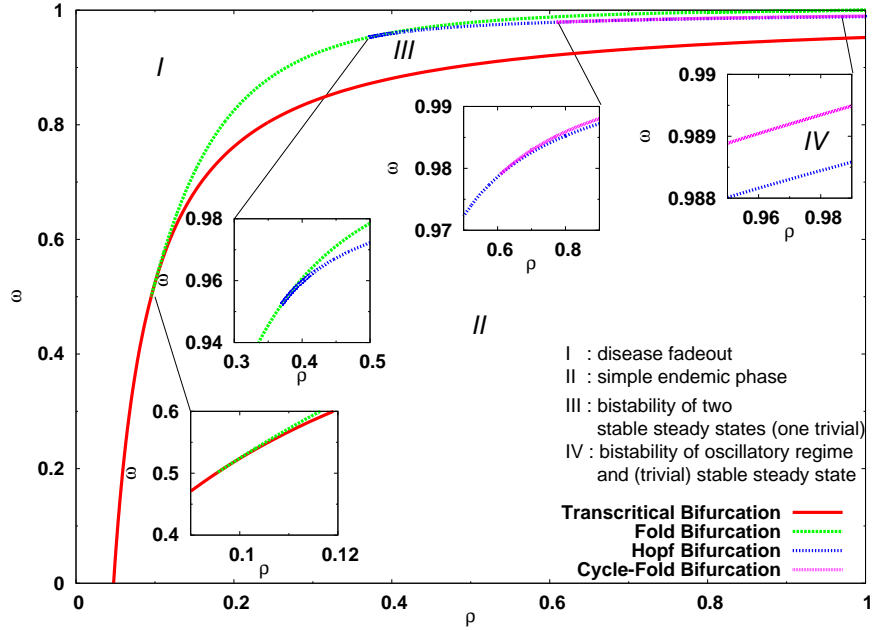}
\caption{Phase diagram of pairwise equations in \cite{Gross} , \(l\!=\!10\). Compiled with \cite{AUTO}.}
\label{phaseplot}
\end{figure}

The full phase diagram is presented in Fig. \ref{phaseplot}; for any \(l\geq0.5\) the minimum rewiring rate for bistability to arise is \(\omega\!=\!0.5\). The transcritical bifurcation, serving as the epidemic threshold and approximated in \cite{Gross}, occurs at exactly \(\omega=1-1/\left(\rho\left[2l+1\right]\right)\). 

We have checked the steady state values for densities and pairs predicted by the PA
against simulations on networks with \(N=10^4\) nodes, taking as 
initial conditions the predicted steady-state value of \(i\), with infected and susceptibles initially
randomly distributed on a Poisson network of average degree \(\langle k\rangle = 2l = 20\).
In the simple endemic phase, prevalence and link densities of numerical simulations and model 
predictions agree reasonably well. At \(\omega=\rho/\left(1+\rho\right)\),
i.e. when the infection rate is equal to the rewiring rate, the pair approximation model correctly predicts \(i\) and the mean degrees of both susceptibles and infected. 
\section{Observing A Steady-State Topology}
\begin{SCfigure}
  \centering
  \includegraphics[width=0.65\textwidth]{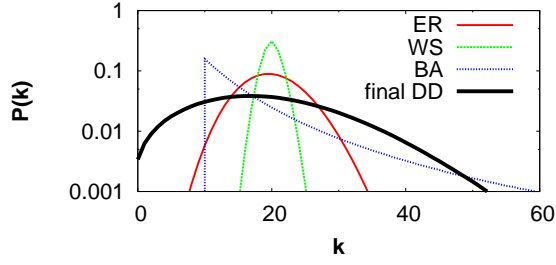}
  \caption{
Initial topologies settling down to final DD. Erd\H{o}s-R\'{e}nyi (ER), Watts-Strogatz with rewiring probability \(\beta=0.1\) (WS), Barabasi-Albert (BA). \(N=10^5,\langle k \rangle=20, \omega=0.9, \rho=0.7\),
averaged over \(10^4\) samples (between \(4000<t<5000\) for final DD).}\label{finalDD}
\end{SCfigure}

\begin{figure}[ht]
  \centering
  \subfloat[][\(\omega=0.1\)]{\label{c1}\includegraphics[width=0.5\textwidth]{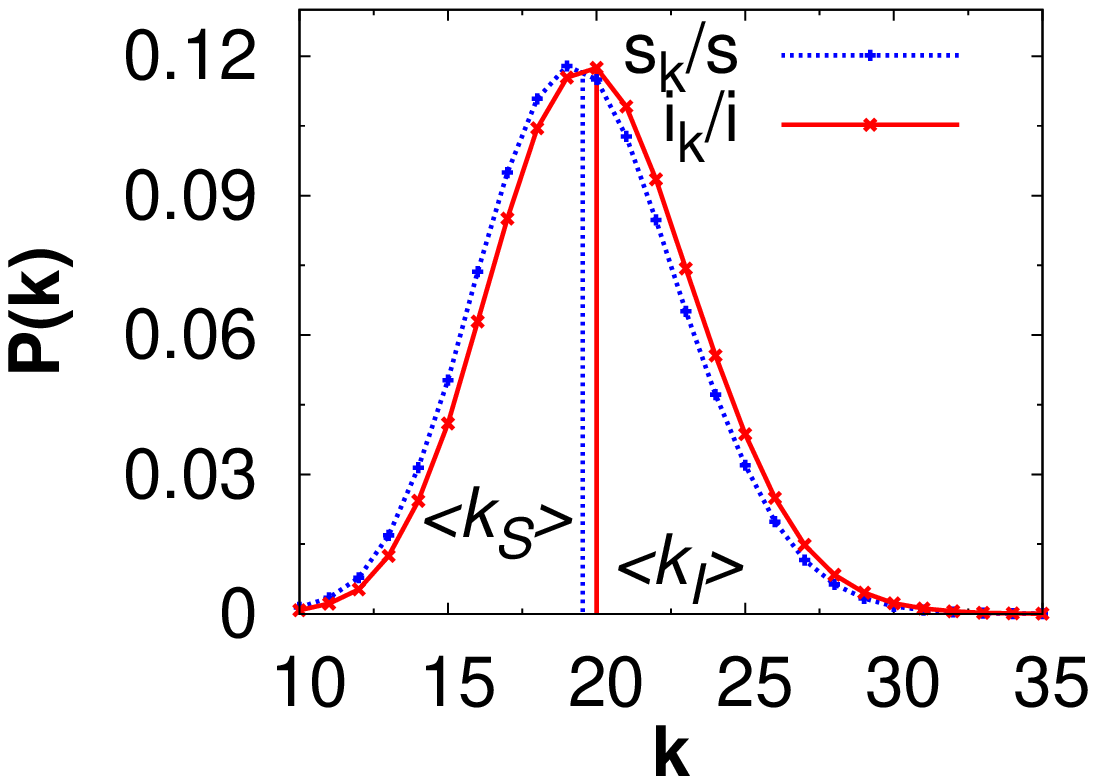}}                
  \subfloat[][\(\omega=0.9\)]{\label{c2}\includegraphics[width=0.5\textwidth]{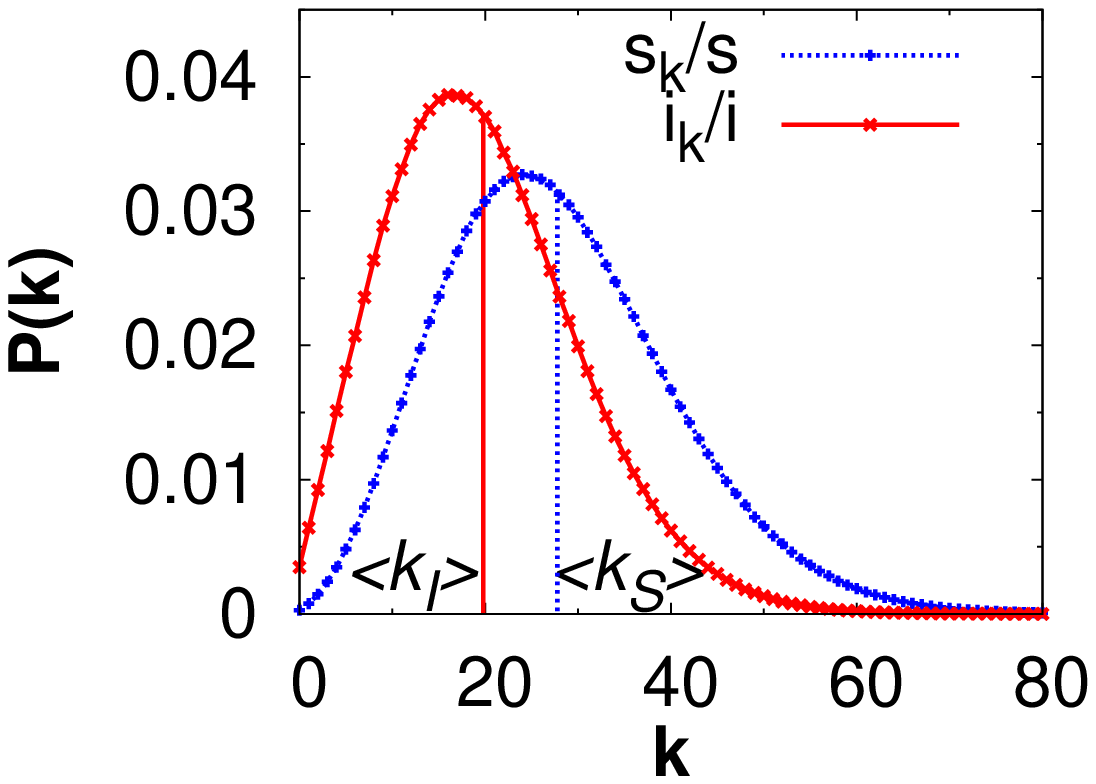}}
  \caption{Infection-status clustering in the steady state in stochastic simulations (\emph{MC}).
 Fig. \ref{c1}: \(\Delta \langle k \rangle_{\mathrm{MC}}=-0.50, \Delta \langle k \rangle_{\mathrm{PA}}=-0.47\).
 Fig. \ref{c2}: \(\Delta \langle k \rangle_{\mathrm{MC}}=8.01, \Delta \langle k \rangle_{\mathrm{PA}}=6.35\).
Initial Poisson DD, \(N=10^5,\langle k \rangle=20, \rho=0.7\), averaged over \(10^4\) samples between \(1000<t<2000\).}
  \label{clustering}
\end{figure}

The steady-state degree distributions (\emph{DD}) of the subnetworks of infected and susceptibles 
- as well as that
of the complete network - are shown in Figs. \ref{finalDD} and \ref{clustering}. 
Independently from topological starting conditions in the simple endemic phase, 
the final topology (as described by several standard measures) settles down 
to a characteristic state (Fig. \ref{finalDD}).

Let \(\langle k_{\mathrm{S}}\rangle = \sum_k k s_k/s\) and \(\langle k_{\mathrm{I}}\rangle = \sum_k k i_k/i\) 
be the average degree of susceptible and infected, where \(a_k\) is the fraction of 
the total number of nodes which are in state \(a\) and have degree \(k\), \(a \in \{s,\enspace i\}\). 
Hence the average degree of the network is \(\langle k\rangle= s \langle k_{\mathrm{S}}\rangle + i 
\langle k_{\mathrm{I}}\rangle\). We have also, by definition, \(s \langle k_{\mathrm{S}}\rangle = 2 l_{\mathrm{SS}} 
+ l_{\mathrm{SI}}\) and \(i \langle k_{\mathrm{I}}\rangle = 2 l_{\mathrm{II}} + l_{\mathrm{SI}}\), so that the average degrees 
of the two subnetworks in the steady state are determined by the equilibrium values
of the PA in the region where the moment closure from \cite{Keeling} holds.

Rewiring induces infection-status clustering (\cite{Gross}); in the PA, the difference 
of the mean degrees of susceptible and infected nodes is independent of the link density 
and given by
\[\Delta \langle k \rangle=\langle k_{\mathrm{S}} \rangle - \langle k_{\mathrm{I}} \rangle= \frac{w}{\left(1-\omega\right)\rho}-1\enspace .\]
Thus only for \(\omega>\rho/\left(1+\rho\right) \) does rewiring yield a bigger mean degree in susceptibles than in infected (Fig. \ref{clustering}). Again \(\omega\) and \(\rho\) given by \(\omega=\rho/\left(1+\rho\right)\) enjoy a special status, in that then the mean degrees of the \(S\)- and \(I\)-ensemble are the same, as are their degree distributions and \emph{every} other topology-related statistics we measured.
\section{The Node Cycle}
Depending on a node's status, it will either gain in degree (when susceptible) or lose links (when infected). Characterising such a typical node cycle gives important single-node characteristics describing its degree during and lifetime of each stage, as well as a means to determine the overall infection-class degree distributions.

The time evolution of the number of susceptible and infected neighbours \(l_{\mathrm{S}}(t)\) and \(l_{\mathrm{I}}(t)\) of a typical node is
\begin{align}\label{phaseI}
l_{\mathrm{S1}}(t) & = c_{\mathrm{1}} e^{-\lambda_{\mathrm{1}} t} +  \lambda_{\mathrm{2}} t+c_{\mathrm{2}}  \nonumber\\
l_{\mathrm{I1}}(t) & = c_{\mathrm{3}} e^{-\lambda_{\mathrm{1}} t} +  \lambda_{\mathrm{3}} t+c_{\mathrm{4}}  
\end{align}
in its susceptible stage (phase I) and
\begin{align}\label{phaseII}
l_{\mathrm{S2}}(t) & = c_5 e^{-\lambda_4 t} + c_6 e^{-\lambda_5 t}\nonumber\\
l_{\mathrm{I2}}(t) & = c_7 e^{-\lambda_4 t} + c_8 e^{-\lambda_5 t}
\end{align}
when it is infected (phase II). The constants \(\lambda_i \in \mathbb{R}_{\geq0}\) 
and \(c_j \in \mathbb{R}\) are comprised of \(\omega, \rho\), the steady-state values of \(l_{\mathrm{SI}}\) and \(s\) (computable through the PA), as well as the four initial conditions to be determined by a cyclic closure of (\ref{phaseI}) and (\ref{phaseII}).

\begin{figure}[ht]
  \centering
  \subfloat[][Model]{\label{cyclemodel}\includegraphics[width=0.5\textwidth]{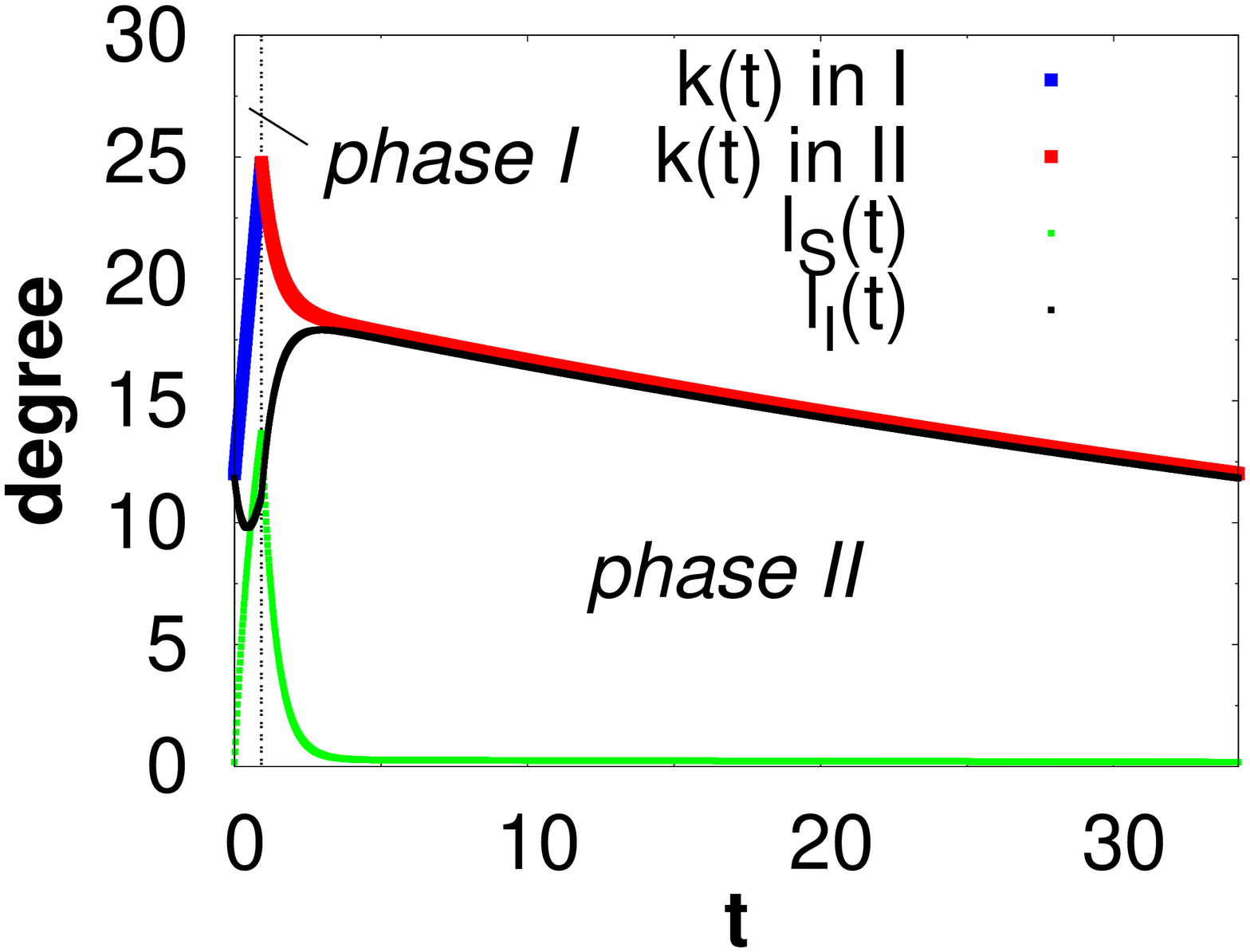}}                
  \subfloat[][Simulations]{\label{cyclesim}\includegraphics[width=0.5\textwidth]{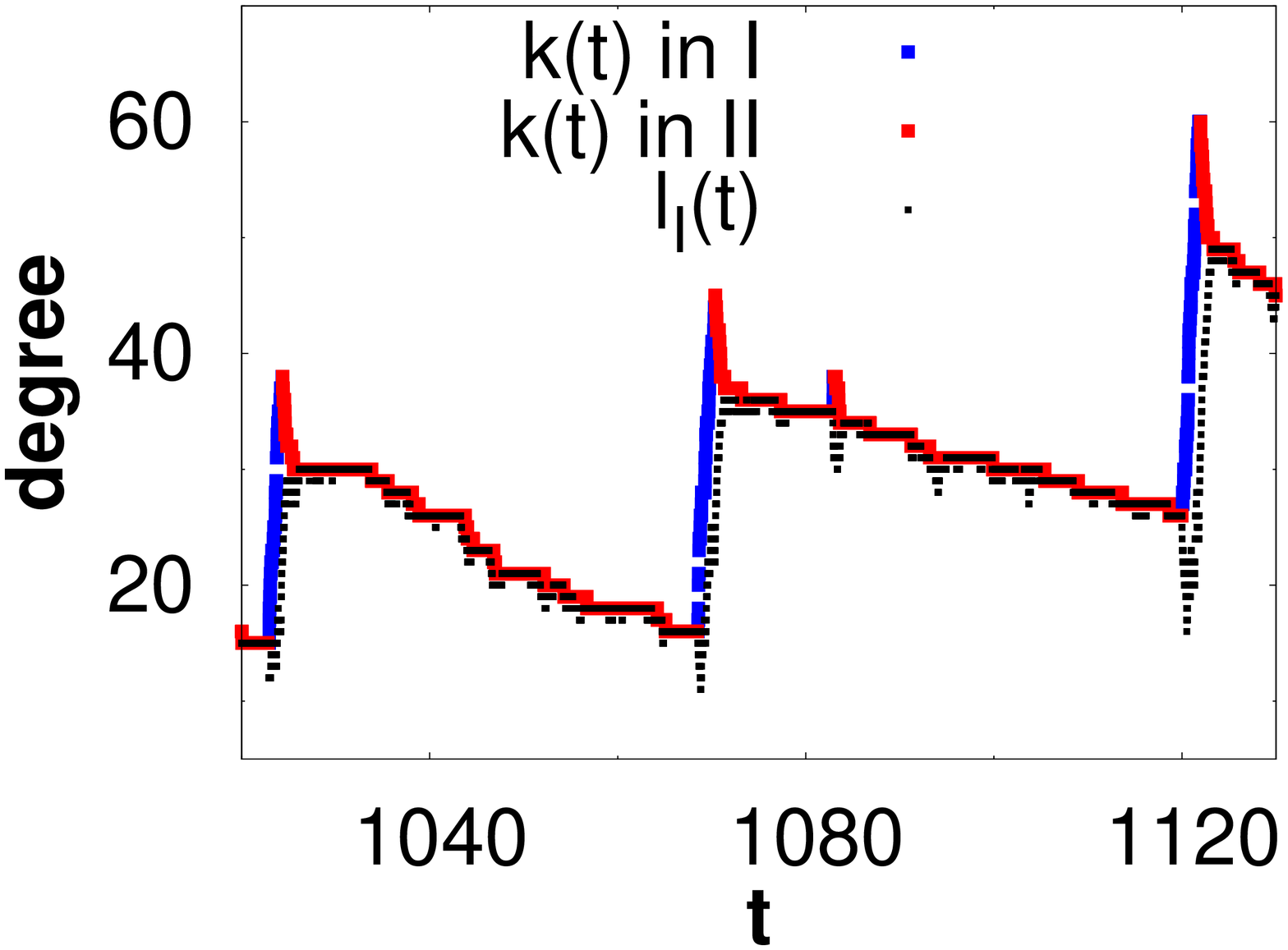}}
\caption{Degree evolution of a single node, \(\omega=0.9,\rho=0.7, l=10\).
 Fig. \ref{cyclemodel}:  \(l_{\mathrm{S1}}(0)=0.43,l_{\mathrm{I1}}(0)=15.6, l_{\mathrm{S2}}(0)=l_{\mathrm{S1}}(\tau_{\mathrm{S}}),l_{\mathrm{I2}}(0)=l_{\mathrm{I1}}(\tau_{\mathrm{S}})\).
 Fig. \ref{cyclesim}: \(N=10^4\).}
  \label{cycle}
\end{figure}

A crucial ingredient is the mean-field approximation for the average lifetime \(\tau_{\mathrm{S}}\) of a susceptible; \(\tau_{\mathrm{S}}=s/\left(\left[1-\omega\right] \rho l_{\mathrm{SI}}\right)\) turns out to be a very good one throughout parameter space. This is due to the fact that over a susceptible's lifetime, there is no noticable degree-, time- or any other correlation in the \(S\)-nodes rewired to it, particularly not in the number of \emph{their} \(I\)-neighbours. Hence the node-cycle model also accurately describes regimes with very high \(\omega\). There, the near-linear degree growth and the trough of \(l_{\mathrm{I}}(t)\) around \(\tau_{\mathrm{S}}\) in the first phase are well-captured by our equations. So is the rapid rewiring away of \(l_{\mathrm{S}}\)-links in the initial stage of the second phase, followed by slow degree loss through recovery and subsequent rewiring (Fig. \ref{cyclemodel} and \ref{cyclesim}).  
\section{Deriving the Degree Distribution}
Let \(l_{\mathrm{SI}_k}\) be the fraction of the \(SI\)-link density with \(I\)-nodes of degree \(k\). 
Denote also by \(l_{\mathrm{S}_k\mathrm{I}}\) the number of \(SI\)-links whose \(S\)-node is of degree \(k\), 
divided by the number of nodes \(N\). In order to describe the infection-class subnetworks, we must, instead of considering the pairwise equations in \cite{Gross}, deal with the dynamics of each degree class of infected and susceptibles. For the processes and 
rates we are considering, this is given by
\begin{align}
 \frac{di_k}{dt}= & - (1-w)(1-\rho) i_k + (1-w)\rho l_{\mathrm{S}_k\mathrm{I}}
+ w 
\left( l_{\mathrm{SI}_{k+1}} - l_{\mathrm{SI}_{k}} \right) \nonumber\\
\frac{ds_k}{dt}= & (1-w)(1-\rho) i_k - (1-w)\rho l_{\mathrm{S}_k\mathrm{I}}
+ w \frac{l_{\mathrm{SI}}}{s}\left( s_{k-1} - s_k  \right) \enspace. 
\label{Deqs0}
\end{align}
\subsection{Inside the PA regime}
Assuming that the end nodes of \(l_{\mathrm{SI}}\)-links uniformly sample the two subpopulations, 
(\ref{Deqs0}) may be closed by taking
\begin{equation}
l_{\mathrm{S}_k\mathrm{I}}\approx  l_{\mathrm{SI}}\frac{k s_k}{s \langle k_{\mathrm{S}}\rangle} \quad \wedge \quad l_{\mathrm{SI}_k}  \approx l_{\mathrm{SI}} \frac{k i_k}{i \langle k_{\mathrm{I}}\rangle} \enspace. 
\label{approx}
\end{equation}
From (\ref{Deqs0}) with \(i\), \(s\), \(\langle k_{\mathrm{S}}\rangle\) and \(\langle k_{\mathrm{I}}\rangle\) in 
(\ref{approx}) given by their values in the steady state of the PA in \cite{Gross}, we obtain recurrence relations for the steady-state values of \(s_k\) and \(i_k\). Together with the 
normalisation conditions, these equations can be solved for the two subnetwork's stationary DD.
In the region where the PA performs well, this is a reasonable approximation (results for 
a sample point in this region are shown in Fig. \ref{DD1}. Also in this region, state-degree 
correlations are negligible, and the DDs of the two subnetworks are practically identical (as observed in simulations before). 

\subsection{Outside the PA regime}
The fully analytic, PA-based procedure described above breaks down whenever the PA does,
and state-degree correlations then become important. Since the PA no longer holds for the 
\(\langle k_{\mathrm{S,I}}\rangle\), these cannot be treated as parameters anymore and must be expressed 
in terms of the variables \(s_k, i_k\) in (\ref{Deqs0}). 
Because of the state-degree correlations, a different closure assumption than (\ref{approx})
must be used for (\ref{Deqs0}). Instead of the second equation in (\ref{approx}),  
we shall take
\begin{align} 
l_{\mathrm{SI}_k}  \approx   l_{\mathrm{SI}} \left( a \frac{k s_k}{s \langle k_{\mathrm{S}}\rangle} + (1-a) \frac{k i_k}
{i \langle k_{\mathrm{I}}\rangle}\right), 
\label{approx2}
\end{align}
where \(a \in [0,1]\) is a phenomenological coefficient obtained through least squares fitting.

The fact that \(a\) must be nonzero can be understood as follows: Links of type \(SI\) are
very short-lived with respect to the average infectious period. This means that the \(I\)-nodes of 
\(SI\)-links do not uniformly sample the infected population, and that the sampling is biased 
towards the 'younger' infected nodes, born from susceptible nodes. This sampling bias can 
then be modelled
as a weighted average of the \(s_k\)- and \(i_k\)-distribution. This argument holds throughout, but
only when state-degree correlations are important, i.e. for \(\omega>\rho/\left(1+\rho\right)\),     
is this correction important. A similar argument applies to the \(l_{\mathrm{S}_k\mathrm{I}}\)-distribution, but the 
sampling bias is far less pronounced in this case, because the average susceptible period is  
smaller than the average infectious period in this region.

\begin{figure}[ht]
  \centering
  \subfloat[][\(\omega=0.5\)]{\label{DD1}\includegraphics[width=0.5\textwidth]{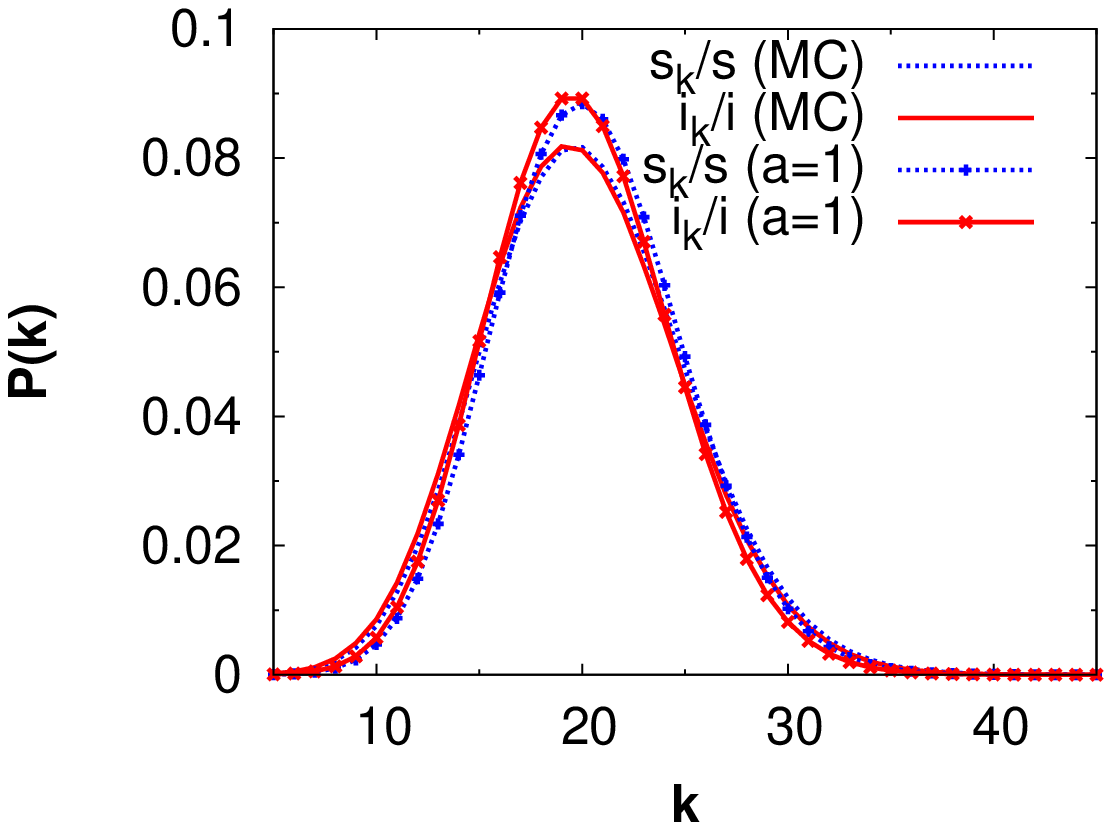}}                
  \subfloat[][\(\omega=0.9\)]{\label{DD2}\includegraphics[width=0.5\textwidth]{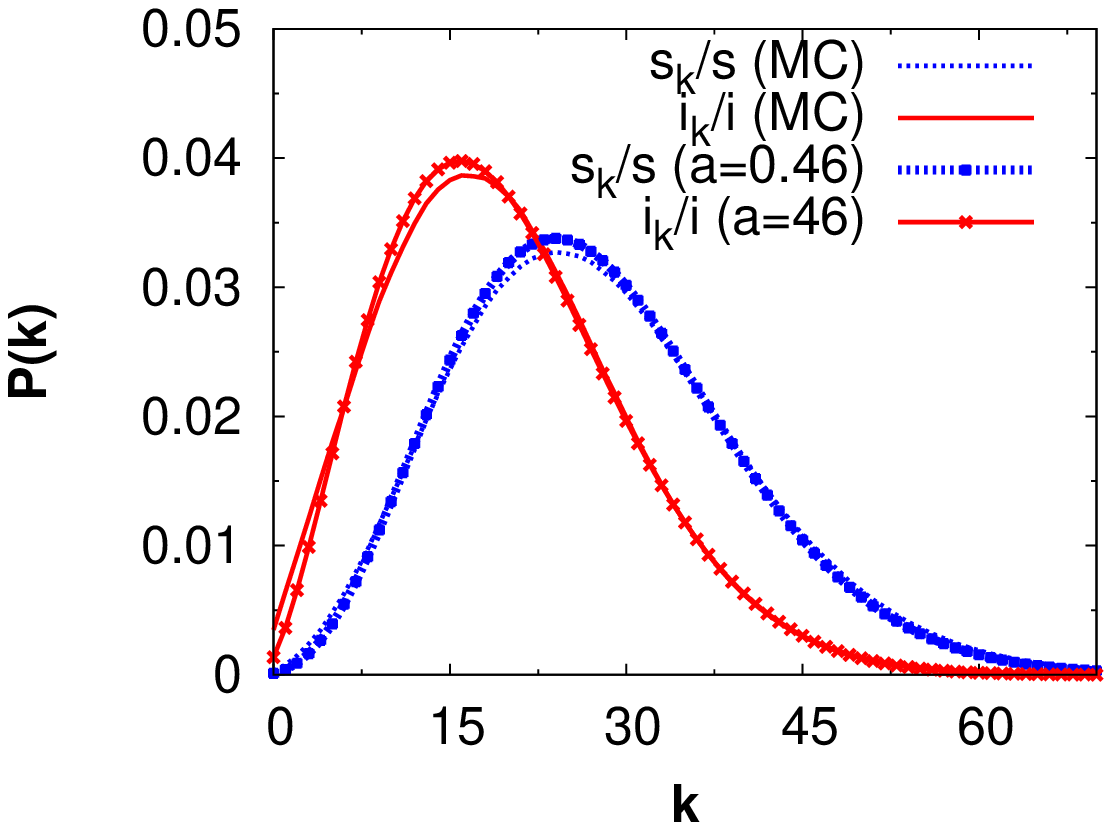}}
\caption{Analytic stationary DD of infected and susceptible subnetworks vs. MC results. Simulations with initial Poisson DD, \(N=10^4,\langle k \rangle=20, \rho=0.7\), averaged over \(10^4\) samples between \(1000<t<2000\).}
  \label{DD}
\end{figure}

With these two modifications with respect to the preceding section,
the equilibrium of (\ref{Deqs0}) can also be found (although no longer through recurrence relations)
and compared with the stationary degree distributions obtained from simulations (Fig. \ref{DD2}). The best choice of the coefficient 
\(a\) gives a good quantitative agreement for the whole range of the DDs; whereas the predictions of 
the naive PA-based approach used before are way off the Monte Carlo results.
\section{Acknowledgements} 
Financial support from the Foundation of the University of Lisbon 
and the Portuguese Foundation for Science and 
Technology (FCT) under contract POCTI/ISFL/2/618 is gratefully 
acknowledged. The first and second authors (SW and TA) were also supported by FCT under grants SFRH/BD/45179/2008 and CFTC-618-BII-02/08. 

\renewcommand\refname{\textbf{References}}
\bibliographystyle{unsrt}
\bibliography{ref}

\end{document}